\documentclass[9pt,shortpaper,twoside,web]{ieeecolor}
\usepackage{generic}
\UseRawInputEncoding
\usepackage{amsmath,amssymb,amsfonts}
\usepackage{graphicx}
\usepackage{textcomp}
\def\BibTeX{{\rm B\kern-.05em{\sc i\kern-.025em b}\kern-.08em
    T\kern-.1667em\lower.7ex\hbox{E}\kern-.125emX}}
\begin{document}
\title{Quantized control of non-Lipschitz nonlinear systems: a novel control framework with prescribed transient performance and lower design complexity}
\author{Zongcheng Liu, Jiangshuai Huang, Changyun Wen, Jing Zhou, Xiucai Huang
\thanks{This work was partially supported by the National Natural Science
Foundation of China under Grant 61603411. The corresponding author is J. S. Huang.}
\thanks{Z. C. Liu is with Aeronautics Engineering College,
Air Force Engineering University, and also with Northwestern Polytechnical University, Xi'an 710038, China (e-mail: liu434853780@163.com).}
\thanks{J. S. Huang and Xiucai Huang are with School of Automation, Chongqing University, Chongqing, China (e-mail: jshuang@cqu.edu.cn, hxiucai@cqu.edu.cn).}
\thanks{C. Y. Wen is with the
School of Electrical and Electronic Engineering, Nanyang Technological
University, Singapore 639798, Singapore (e-mail: ecywen@ntu.edu.sg)}
\thanks{J. Zhou is with the Department of Engineering Sciences, University of
Agder, Grimstad 4898, Norway (e-mail: jing.zhou@uia.no)}
}

\maketitle

\begin{abstract}
A novel control design framework is proposed for a class of non-Lipschitz nonlinear systems with quantized states, meanwhile prescribed transient performance and lower control design complexity could be guaranteed. Firstly, different from all existing control methods for systems with state quantization, global stability of strict-feedback nonlinear systems is achieved without requiring the condition that the nonlinearities of the system model satisfy global Lipschitz continuity. Secondly, a novel barrier function-free prescribed performance control (BFPPC) method is proposed, which can guarantee prescribed transient performance under quantized states. Thirdly, a new \textit{W}-function-based control scheme is designed such that virtual control signals are not required to be differentiated repeatedly and the controller could be designed in a simple way, which guarantees global stability and lower design complexity compared with traditional dynamic surface control (DSC). Simulation results demonstrate the effectiveness of our method.
\end{abstract}


\section{Introduction}
\label{sec:introduction}
In the past decades, controlling uncertain nonlinear systems has always been a hot research topic in the control community [1]-[6]. Among all the control problems of uncertain nonlinear systems, two of them are particularly interested, i.e., how to achieve better control performance and how to reduce the complexity of control design. For the former problem, a significant progress has been made in [7], in which prescribed performance has been achieved by combining barrier functions such that the output error converges to a predefined arbitrarily small residual set, with convergence rate no less than a given prescribed value, and maximum overshoot less than a preassigned level. Inspired by this barrier functions-based control design method, many remarkable results have been obtained [8]-[12], [20]. Though many theoretical problems have been successfully solved using PPC, there are still some problems that should not be neglected, yet remain unsolved. The mechanism of PPC method is actually based on the fact that the barrier functions in controller will approach infinite when the designed error approaches to the barrier value, and thus the controller will always have large enough actions responding to the increasing error. However, the barrier functions will bring some problems for a control system. Firstly, just as pointed out in [13], the system signals must be continuous, since the discontinuity will make the barrier functions-based controller singular. Secondly, high precision measurement of system states is necessary for PPC method, since low precision measurement may include some noises which results in discontinuity. Unfortunately, it is always hard to ensure high precision measurement to satisfy the requirement of barrier functions in practical systems.

On the other aspect, it is well known that backstepping control method faces the problem of ``explosion of complexity'' resulting from the repeated differentiation of virtual controls. Dynamic surface control (DSC) method [14] was therefore designed to solve this problem, where a compact set is introduced and it is finally proved to be an invariant set. DSC method has been widely used for many control problems [14]-[19], since it significantly reduces the complexity of control design. However, only semi-global results can be achieved since some constraints on initial conditions should be strictly satisfied in DSC. The parameter tuning of the DSC method is also time-consuming and there is no guideline in general.

With the digital implementation of control algorithms and development of networked control systems, signal quantization widely exists in practical systems [21]-[23], which is owing to the widespread use of digital processors that employ a finite-precision arithmetic, meanwhile it requires less communication resources. Progress has been made on the quantized control of uncertain systems with input or state quantization. For a control system with state quantization, the state measurements are processed by quantizer, which are discontinuous maps from continuous spaces to finite sets. The discontinuous property will make the control design and stability analysis difficulty, and it will lead to the failure of barrier functions-based control method. A novel adaptive backstepping control is proposed for quantized systems with matched uncertainties in [24]. Employing neural networks [25]-[26] or command filters [27], semi-global control schemes have been designed for systems with quantized states. However, it is worth mentioning that, all existing results of state quantization require that system nonlinearities always satisfy global Lipschitz continuity condition so as to achieve global stability. Therefore, how to achieve or ensure global control with quantized states for nonlinear systems without global Lipschitz continuity condition is still an unsolved problem.

Inspired by the aforementioned problems, this paper aims to propose a low-complexity prescribed performance control method without using barrier functions, and apply this method to solve the outstanding problem for uncertain nonlinear systems under state quantization, with prescribed control performance being guaranteed. It is worthy emphasizing that no existing control method is available for this problem. To achieve the control objective, two main challenges encountered must be tackled. 1) Since discontinuous state quantization is applied, the traditional barrier functions are not applicable. Therefore new control scheme to guarantee transient performance must be proposed; 2) Without global  Lipschitz continuity for the system nonlinearities, how to solve the global control with state quantization remains unknown and thus is tricky. To address these issues, we propose a novel control method for uncertain nonlinear systems with guaranteed transient performance. The main contributions of this paper are summarized as follows:

1) Without using barrier functions, a novel barrier function-free prescribed performance control (BFPPC) method is proposed by constructing an invariant set and adequately using the upper bounds of states and errors on the invariant set. Comparing with PPC method, the BFPPC method allows controller to be designed by using discontinuous quantized system states. This enables the BFPPC to have broader application potential than PPC in practical systems as discontinuous states can be considered.

2) Comparing with all the existing results on nonlinear systems with quantized states, two obvious achievements are made. Firstly, global boundedness of strict-feedback nonlinear systems with quantized states is achieved, without requiring system nonlinearities to satisfy the global Lipschitz continuity conditions. Secondly, prescribed performance for system output and states is guaranteed.

3) Combining with the properties of $W$-functions, the BFPPC controller is low-complex in the sense that virtual control signal is only the function of state error, and it does not need to be differentiated repeatedly in the controller design. Compared with the DSC-based method, global stability is achieved and an explicit guideline for tuning the control parameters is provided.

\section{BFPPC with quantized states}
In this section, we will show the BFPPC design for uncertain nonlinear
systems with quantized states in order to show the advantage of BFPPC
comparing to PPC methods.
\subsection{Problem statement and preliminaries}
Consider the following system
\begin{equation}
\left\{ {\begin{array}{l}
 \dot {x}_i =f_i (\bar {x}_i )+x_{i+1} ,\quad i=1,...,n-1 \\
 \dot {x}_n =f_n (\bar {x}_n )+u \\y=x_1
 \end{array}} \right.
\end{equation}
where $\bar {x}_i =[x_1 ,x_2 ,...,x_i ]^T\in R^i$ denotes the state vector
of the system; $u\in R$ is system control input; $y\in R$ is system output;
$f_i (\cdot )$ are uncertain system nonlinearities,
$i=1,...,n$.

\textit{Assumption 1}: $f_i (\cdot ), i=1,...,n,$ is a continuous function
satisfying $\left| {f_i (x_1 ,...,x_i )} \right|\le f_i^\ast (x_1 ,...,x_i
)$ where
$f_i^\ast (\cdot )$ is known function.

\textit{Assumption 2 [24]}: Only quantized states $\left( {q(x_1 ),q(x_2 ),...,q(x_n
)} \right)$ are measurable and available for control design, instead of the
states $\left( {x_1 ,x_2 ,...,x_n } \right)$.

The quantizer $q(x)$ considered in this paper can be either of uniform
quantizer, hysteresis-uniform quantizer or logarithmic-uniform quantizer
mentioned in [24], where the quantizer $q(x)$ has the following property
\begin{equation}
\left| {q(x)-x} \right|\le \delta _0 ,\quad \forall x\in R
\end{equation}
where $\delta _0 >0$ is a known constant representing the quantization
bound. The map of the uniform quantizer $q(x)$ for $x>0$ is shown in Fig. 1.
\begin{figure}[h!]
    \includegraphics[width=2.6in,height=2.6in,clip,keepaspectratio]{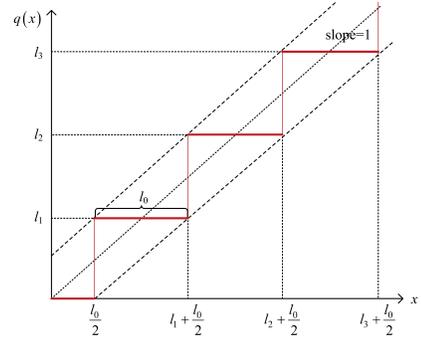}\\
  \raggedleft
  \caption{Map of uniform quantizer $q(x)$}
\end{figure}

A uniform quantizer can be modeled as
\begin{equation}
q(x)=\left\{ {\begin{array}{l}
 l_i \quad l_i -\frac{l_0 }{2}\le x<l_i +\frac{l_0 }{2} \\
 0\quad -\frac{l_0 }{2}\le x<\frac{l_0 }{2} \\
 -l_i \quad -l_i -\frac{l_0 }{2}\le x<-l_i +\frac{l_0 }{2} \\
 \end{array}} \right.\quad
\end{equation}
where $l_{i+1} =l_i +\frac{l_0 }{2}, i=0,1,2,...$, and $l_0 $ is the length of the
quantization interval. $q(x)$ is in the set $U=\{0,\pm l_i \}$. The
quantization error is bounded by (2), where $\delta _0 \ge \frac{l_0 }{2}$.
From Fig. 1 and (3), it can be seen that the quantized states $\left(
{q(x_1 ),q(x_2 ),...,q(x_n )} \right)$ for control design are not continuous
signals, which will make the barrier function-based methods fail since
PPC or other barrier function-based methods require the system states to be
continuous signals.  Now we define the global control in the sense of prescribed performance as follows.

\textit{\textbf{Global barrier-function-free prescribed performance control (GBPPC) problem}}:
Consider a class of uncertain nonlinear systems. Without using barrier functions, design a controller $u$ such
that for any initial conditions $x_i (0)=x_i^0 $, $i=1,...,n$, all signals in
the closed-loop system remain bounded and the system output is
confined to prescribed area with prescribed convergence rate and maximum
overshoot.

The control objective of this section is to solve GBPPC problem of system (1) with quantized states. To this end, the $W$-function is proposed as follows, which is the key to achieve the control objective.

\textit{Definition 1 ($W$-function)}: Let $k$  be any positive integer and define a
continuous differentiable function $F(z_1 ,z_2 ,...,z_k )$: $R^+\times R^+\times \cdots \times R^+\to [0,+\infty )$, where
$R^+=[0,+\infty )$. Then, $F(\cdot )$ is called a $W$-function, if
$\frac{\partial F(z_1 ,z_2 ,...,z_k )}{\partial z_j }\ge 0$ and $F(z_1 ,z_2
,...,z_k )>0$ for $\forall z_j \in R^+,j=1,...,k$.

\textit{Lemma 1 [28] (Separation theorem)}: For any real-valued continuous function
$f(x,y)$, where $x\in R^m$, $y\in R^n$, there are smooth scalar functions
$a(x)\ge 0$, $b(y)\ge 0$, $c(x)\ge 1$ and $d(y)\ge 1$ such that
\begin{equation}
\left| {f(x,y)} \right|\le a(x)+b(y)
\end{equation}
\begin{equation}
\left| {f(x,y)} \right|\le c(x)d(y)
\end{equation}

\textit{Remark 1}: For an arbitrary continuous function $f(\cdot )$, it is easy to
find a $W$-function $F(\cdot )$ such that $\left| {f(\cdot )} \right|\le
F(\cdot )$, as illustrated in the following examples, 1) if $f_2(x_1 ,x_2 )=x_1^2 +x_1 \sin x_1 +x_1 x_2 $,
then, the $W$-function $F_2 (|x_1| ,|x_2| )=x_1^2 +\left| {x_1 } \right|+\left|
{x_1 x_2 } \right|$ satisfies $\left| {f_2(x_1 ,x_2 )} \right|\le F_2(|x_1| ,|x_2|
)$, 2) if $f_2(x_1 ,x_2 )=x_2 x_1^2 +e^{x_2 }\cos x_2 $, then, the $W$-function
$F_2 (|x_1| ,|x_2| )=\left| {x_2 } \right|x_1^2 +e^{\left| {x_2 } \right|}$
satisfies $\left| {f_2(x_1 ,x_2 )} \right|\le F_2(|x_1| ,|x_2| )$.

\textit{Remark 2}: It should be noted that, though there are many different methods that achieved prescribed performance for tracking error, such as low-complexity control [12], [20], prescribed performance control [7], [8], [10], and barrier function-based control [9]. It should be noted that, they are all barrier functions-based methods because barrier functions is critical and essential in their methods in confining the tracking error within the prescribed performance functions. However, we want to strengthen that all the control schemes mentioned above work based on the fact that the control signals approach infinity if some states approach the pre-defined ``barrier'' and all the control schemes mentioned above fail to work in discrete control environment such as state quantization. For example, for the PPC-based control scheme proposed in [7] in face of state quantization, the matrix R in (8) of [7] may be singular. For the barrier Lyapunov function-based control, this problem is more obvious. Therefore the PPC-based control scheme in [7] is also called ``Barrier Lyapunov function-based control'', and it is not applicable in discrete case such as state quantization. This paper will give a prescribed performance control to solve this problem without using barrier function for the first time.

\subsection{Controller design}
In view of the quantized states, introduce the following
change of coordinates
\begin{equation}
e_1^{(q)} =q(x_1 (t))-\rho (t)x_1^{(q)} (0)
\end{equation}
\begin{equation}
e_i^{(q)} =q(x_i (t))-\alpha _{i-1}^{(q)} -\rho (t)x_i^{(q)} (0),i=2,...,n
\end{equation}
where $x_i^{(q)} (0)=q\left( {x_i (0)} \right),i=1,...,n$, and $\rho (t)$ is
a performance function. For the sake of brevity, in this section, choose
\begin{equation}
\rho (t)\mbox{=}\left\{ {\begin{array}{l}
 \frac{1}{2}\left( {1+\cos \left( {\frac{t}{t_s }} \right)}
\right),\quad t<\pi t_s \\
 0,\quad t\ge \pi t_s \\
 \end{array}} \right.
\end{equation}
where $t_s $ is an arbitrary positive constant. It can be easily seen that $0\le {\rho (t)}  \le 1$ and $\left| {\dot \rho (t)} \right| \le {\rho _M}$ with ${\rho _M} > 0$ being a known constant. More details on the design of $\rho
(t)$ will be shown in Section III. Then, the virtual and actual controller
are designed as follows
\begin{equation}
\alpha _i^{(q)} =-\gamma _i H_i (\rho (t))e_i^{(q)} -c_i \left[ {e_i^{(q)} }
\right]^{N_i },i=1,...,n
\end{equation}
\begin{equation}
u=\alpha _n^{(q)}
\end{equation}
where $\gamma _i $ and $c_i $ are positive constants, $N_i >1$ is an arbitrary odd
 integer, and $H_i (\rho (t))$ is a design function chosen as (25), (31) and (37) in the following part.

\subsection{Main results}
For the designed control method, we have the following results.

\textbf{\textit{Theorem 1}}: Consider the closed loop system consisting of system (1) and the controller proposed as in (9) and (10). Then, under Assumption 1 and 2, there exist $\gamma _i $, $c_i $, $N_i$ and
$H_i (\rho (t))$ such that:

1). The GBPPC problem of system (1) is solved.

2). The system output satisfies the prescribed performance. Specially, $\rho (t)x_1 (0)-\delta _M -\varepsilon _1 \le x_1 (t)\le \rho
(t)x_1 (0)+\delta _M +\varepsilon _1 $, where $\delta _M $ is a constant
satisfying $\delta _M \ge (1+\rho (t))\delta _0 $ and $\varepsilon _1 $ is
an arbitrary small positive constant.

\textit{Proof}: Since it is difficult to analyze system stability with quantized
errors $e_i^{(q)} $, to facilitate the system stability analysis,
define
\begin{equation}
e_1 =x_1 (t)-\rho (t)x_1 (0),
\end{equation}
\begin{equation}
e_i =x_i (t)-\alpha _{i-1} -\rho (t)x_i (0),i=2,...,n,
\end{equation}
\begin{equation}
\alpha _i (t)=-\gamma _i H_i (\rho (t))e_i -c_i e_i^{N_i } ,i=1,...,n,
\end{equation}
and a compact set
\begin{equation}
\Omega =\left\{ {(e_1 ,...,e_n )\left| {\left| {e_i } \right|\le p_i }
\right.(t),i=1,...,n} \right\}
\end{equation}
where $p_1 (t)=\delta _M +\varepsilon _1 $, $p_i (t)=\delta _{i-1}
(t)+\delta _M +\varepsilon _i $, $\delta _1 (t)=\gamma _1 H_1 (\rho
(t))\delta _M +c_0 $, $\delta _i (t)=\gamma _i H_i (\rho (t))\left( {\delta
_M +\delta _{i-1} (t)} \right)+c_0 $, $i=2,...,n$, with $c_0 $ and
$\varepsilon _i $ being arbitrary positive constants. $H_i (\rho (t))$ will be designed as continuous functions, and $H_i (\rho (t))$ are bounded
by noting $0\le \rho (t)\le 1$, and thus $\delta _i (t)$ and $p_i (t)$
are bounded.

Denote $e=[e_1 ,...,e_n ]^T$. Then, we will prove that $e(t)\in \Omega $ for
$\forall t\ge 0$, i.e., $\Omega $ is an invariant set for $e(t)$. It is
easily verified that $e(0)\in \Omega $ since $e_i (0)=e_i^{(q)} (0)=0$ for
$i=1,...,n$. Then we prove the results through the following n recursive steps.

Step 1: Consider the following Lyapunov function
\begin{equation}
V_1 =\frac{1}{2}e_1^2
\end{equation}

It follows from (1) and (11) that the time derivative of $V_1 $ is
\begin{equation}
\begin{array}{l}
 \dot {V}_1 =e_1 \dot {e}_1 =e_1 (f_1 (\bar {x}_1 )+x_2 -\dot {\rho }(t)x_1
(0)) \\
 =e_1 f_1 (\bar {x}_1 )-e_1 \dot {\rho }(t)x_1 (0)+e_1 (\alpha _1 +\rho
(t)x_2 (0)+e_2 ) \\
 \end{array}
\end{equation}

Noting $x_1 =e_1 +\rho(t)x_1 (0)$, it follows from Assumption 1 that
\begin{equation}
\left| {f_1 (x_1 )} \right|\le f_1^\ast (x_1 )=\phi _{1,0} (\rho(t),e_1
,x_1 (0))
\end{equation}
where $\phi _{1,0} (\cdot )$ is a known function. Obviously, there exists a
$W$-function $\phi _{1,1} (\cdot )$ such that
\begin{equation}
\phi _{1,0} (\rho(t),e_1 ,x_1 (0))\le \phi _{1,1} (\rho(t),\left|
{e_1 } \right|,\left| {x_1 (0)} \right|)
\end{equation}
which means
\begin{equation}
\left| {f_1 (x_1 )} \right|\le \phi _{1,1} (\rho(t),\left| {e_1 }
\right|,\left| {x_1 (0)} \right|)
\end{equation}

Then, there exists a $W$-function $h_1 (\cdot )$ such that
\begin{equation}
\begin{array}{l}
 h_1 (\rho (t),\left| {\dot {\rho }(t)} \right|,\left| {e_1 } \right|,\left|
{x_1 (0)} \right|,\left| {x_2 (0)} \right|,\delta _0 ) \\
 \ge \left| {f_1 (\bar {x}_1 )-\dot {\rho }(t)x_1 (0)+\rho(t)x_2 (0)}
\right| \\
 \end{array}
\end{equation}

Invoking $\left| {x_1 (0)-x_1^{(q)} (0)} \right|\le \delta _0 $ and $\left|
{x_2 (0)-x_2^{(q)} (0)} \right|\le \delta _0 $, $x_1 (0)$ and $x_2 (0)$ can
be expressed as
\begin{equation}
x_1 (0)=x_1^{(q)} (0)+\theta _1 \delta _0
\end{equation}
\begin{equation}
x_2 (0)=x_2^{(q)} (0)+\theta _2 \delta _0
\end{equation}
with constants $-1\le \theta _1 \le 1$, $-1\le \theta _2 \le 1$.

Using Lemma 1 and (21)-(22), there exists $W$-functions $h_2 (\cdot )$ and $h_3 (\cdot )$ such
that
\begin{equation}
\begin{array}{l}
 h_2 (\rho (t),\left| {\dot {\rho }(t)} \right|,\left| {e_1 } \right|,\left|
{x_1^{(q)} (0)} \right|,\left| {x_2^{(q)} (0)} \right|,\delta _0 )  +h_3 (|\theta _1| \delta _0 ,|\theta _2| \delta _0 ) \\
 \ge h_1 (\rho (t),\left| {\dot {\rho }(t)} \right|,\left| {e_1 }
\right|,\left| {x_1 (0)} \right|,\left| {x_2 (0)} \right|,\delta _0 ) \\
 \end{array}
\end{equation}

Then, choose $H_1^0 (\cdot)$ and $H_1 (\rho (t))$ as
\begin{equation}
\begin{array}{l}
 H_1^0 (\rho (t),\left| {\dot {\rho }(t)} \right|,\left| {e_1 }
\right|,\left| {x_1^{(q)} (0)} \right|,\left| {x_2^{(q)} (0)} \right|,\delta
_0 )=h_3 (\delta _0 ,\delta _0 ) \\
~~~~~~~~~~~~~~~~~+h_2 (\rho (t),\left| {\dot {\rho }(t)} \right|,\left| {e_1 }
\right|,\left| {x_1^{(q)} (0)} \right|,\left| {x_2^{(q)} (0)} \right|,\delta
_0 )\\
 \end{array}
\end{equation}
\begin{equation}
H_1 (\rho (t))=H_1^0 (\rho (t),\rho _M ,p_1 ,\left| {x_1^{(q)} (0)}
\right|,\left| {x_2^{(q)} (0)} \right|,\delta _0 )
\end{equation}

Using (20) and (23)-(25), we have
\begin{equation}
H_1 (\rho (t))\ge | {f_1 (\bar {x}_1 )} - {\dot {\rho }
(t)x_1 (0)} + {\rho(t)x_2 (0)} |
\end{equation}
on $\Omega $.

Noting $\left| {e_2 } \right|\le p_2 $ on $\Omega $, substituting (13) and
(26) into (16) yields
\begin{equation}
\begin{array}{l}
 \dot {V}_1 =-\gamma _1 H_1 (\rho (t))e_1^2 -c_1 e_1^{N_1 +1} +e_1 e_2 \\
 ~~~~~~~~+e_1 (f_1 (\bar {x}_1 )- \dot {\rho }(t)x_1 (0)+\rho (t)x_2 (0)) \\
 ~~~\le \left| {e_1 } \right|\left( {H_1 (\rho (t))-\gamma _1 H_1 (\rho
(t))\left| {e_1 } \right|-c_1 \left| {e_1 } \right|^{N_1 }+p_2 } \right) \\
 \end{array}
\end{equation}

Step $i\;(i=2,...,n-1)$: Consider the following Lyapunov function
\begin{equation}
V_i =\frac{1}{2}e_i^2
\end{equation}

It follows from (1) and (12) that the time derivative of $V_i $ is
\begin{equation}
\begin{array}{l}
 \dot {V}_i =e_i \dot {e}_i =e_i (f_i (\bar {x}_i )+x_{i+1} -\dot {\rho
}(t)x_i (0)-\dot {\alpha }_{i-1} ) \\
 =e_i f_i (\bar {x}_i )+e_i (\alpha _i+\rho (t)x_{i+1} (0)+e_{i+1}-\dot {\rho }(t)x_i (0)-\dot {\alpha }_{i-1}
) \\
 \end{array}
\end{equation}

Similar to Step 1, there exists a $W$-function $H_i^0 (\cdot )$ such that
\begin{equation}
\begin{array}{l}
 H_i^0 (\rho (t),\left| {\dot {\rho }(t)} \right|,\left| {e_1 }
\right|,...,\left| {e_i } \right|,\left| {x_1^{(q)} (0)} \right|,...,\left|
{x_{i+1}^{(q)} (0)} \right|,\delta _0 ) \\
 \ge \left| {f_i (\bar {x}_i )-\dot {\rho }(t)x_i (0)-\dot {\alpha }_{i-1}
+\rho (t)x_{i+1} (0)} \right| \\
 \end{array}
\end{equation}

Therefore, choose
\begin{equation}
H_i (\rho (t))=H_i^0 \Big(\rho (t),\rho _M ,p_1 ,...,p_i ,\left| {x_1^{(q)} (0)}
\right|,...,\left| {x_{i+1}^{(q)} (0)} \right|,\delta _0 \Big)
\end{equation}

Using (30), (31) and the property of $W$-function, we have
\begin{equation}
H_i (\rho (t))\ge \left| {f_i (\bar {x}_i )-\dot {\rho }(t)x_i (0)-\dot {\alpha }_{i-1}
+\rho (t)x_{i+1} (0)} \right|
\end{equation}
on $\Omega $.

Noting $\left| {e_{i+1} } \right|\le p_{i+1} $ on $\Omega $, substituting
(13) and (32) into (29) yields
\begin{equation}
\begin{array}{l}
 \dot {V}_i =-\gamma _i H_i (\rho (t))e_i^2 -c_i e_i^{N_i +1} +e_i e_{i+1}
\\
 ~~~~~~+e_i (f_i (\bar {x}_i )- \dot {\rho }(t)x_i (0)-\dot {\alpha }_{i-1}+\rho (t)x_{i+1} (0)) \\
  ~~~\le \left| {e_i } \right|\left( {H_i (\rho (t))-\gamma _i H_i (\rho
(t))\left| {e_i } \right|-c_i \left| {e_i } \right|^{N_i }+p_{i+1} } \right)
\\
 \end{array}
\end{equation}

Step $n$: Consider the following Lyapunov function
\begin{equation}
V_n =\frac{1}{2}e_n^2
\end{equation}

It follows from (1), (10) and (13) that the time derivative of $V_i $ is
\begin{equation}
\begin{array}{l}
 \dot {V}_n =e_n \dot {e}_n =e_n (f_n (\bar {x}_n )+u-\dot {\alpha }_{n-1}-\dot {\rho }(t)x_n
(0)) \\
 =e_n (f_n (\bar {x}_n )-\dot {\alpha }_{n-1}- \dot {\rho }(t)x_n (0))+e_n (\alpha _n +\alpha
_n^{(q)} -\alpha _n ) \\
 =-\gamma _n H_n (\rho (t))e_n^2 -c_n e_n^{N_n +1} +e_n (\alpha _n^{(q)}
-\alpha _n ) \\
 ~~~+e_n (f_n (\bar {x}_n )-\dot {\alpha }_{n-1}- \dot {\rho }(t)x_n (0)) \\
 \end{array}
\end{equation}

Similar to the former steps, there exists a $W$-function $H_n^0 (\cdot )$
such that
\begin{equation}
\begin{array}{l}
 H_n^0 (\rho (t),\left| {\dot {\rho }(t)} \right|,\left| {e_1 }
\right|,...,\left| {e_n } \right|,\left| {x_1^{(q)} (0)} \right|,...,\left|
{x_n^{(q)} (0)} \right|,\delta _0 ) \\
 \ge \left| {f_n (\bar {x}_n )-\dot {\alpha
}_{n-1} }- \dot {\rho }(t)x_n (0) \right| \\
 \end{array}
\end{equation}

Therefore, choose
\begin{equation}
H_n (\rho (t))=H_n^0 \Big(\rho (t),\rho _M ,p_1 ,...,p_n ,\left| {x_1^{(q)} (0)}
\right|,...,\!\left| {x_n^{(q)} (0)} \right|\!,\!\delta _0 \Big)
\end{equation}

Then, we have
\begin{equation}
\dot {V}_n\! \le \!\left| {e_n } \right|\left( {H_n (\rho (t))\!-\!\gamma _n H_n
(\rho (t))\left| {e_n } \right|\!\!-\!c_n \left| {e_n } \right|^{N_n }\!+\!\left|
{\alpha _n^{(q)} \!-\!\!\alpha _n } \right|} \right)
\end{equation}
on $\Omega $.

Choose parameters $c_i >0$ and $\gamma _i >0$, $i=1,...,n$, to satisfy
\begin{equation}
\left\{ {\begin{array}{l}
 0 < {c_1} \le {c_0}{\left( {{N_1}{\delta _M}\left( {p_1^{{N_1} - 1} + {{\left( {{p_1} + {\delta _M}} \right)}^{{N_1} - 1}}} \right)} \right)^{ - 1}} \\
 0\!<\!c_i\! \le\! c_0 \left( {N_i \left( {\delta _M \!+\!\delta _{i-1} } \right)\left(
{p_i^{N_i -1} \!\!+\!\!\left(p_i\!+ \!{\delta _M\! +\!\delta _{i-1} } \right)^{N_i -1}} \right)}
\right)^{-1} \\
 i=2,...,n \\
 \gamma _i >\frac{H_i^\ast +\varepsilon _{i+1} +\delta _M+c_0 -c_i p_i^{N_i }
}{\varepsilon _i H_i^\ast },~~~
 i=1,...,n-1\\
 \gamma _n >\!\frac{H_n^\ast\! +c_0 -c_n p_n^{N_n }
}{\varepsilon _n H_n^\ast }
\\
 \end{array}} \right.
\end{equation}
where $H_i^\ast =H_i^0 (0,...,0)$ are positive constants $i=1,...,n$. It
should be noted that $H_i^\ast $ is a positive constant, since
$H_i^0 (\cdot )$ is designed as the upper bound functions for some uncertain
functions and thus we can always increase $H_i^0 (\cdot )$ to avoid
$H_i^\ast =0$.

 Noting (6), (11), the first inequality of (39), using the first inequality of Lemma 3 in [29] and employing $|e_1|\le p_1$ on $\Omega $, we have
\begin{equation}
\begin{array}{l}
\left| {{c_1}\left( {e_1^{{N_1}} - {{\left[ {e_1^{(q)}} \right]}^{{N_1}}}} \right)} \right|\\
 \le \left| {{c_1}{N_1}\left( {{e_1} - e_1^{(q)}} \right)\left( {e_1^{{N_1-1}} + {{\left[ {e_1^{(q)}} \right]}^{{N_1-1}}}} \right)} \right|\\
 \le \left| {{c_1}{N_1}{\delta _M}\left( {p_1^{{N_1} - 1} + {{\left( {{p_1} + {\delta _M}} \right)}^{{N_1} - 1}}} \right)} \right| \le {c_0},
\end{array}
\end{equation}

By noting (9), (13) and (40), we obtain
\begin{equation}\begin{array}{l}
\left| {\alpha _1^{(q)} -\alpha _1 } \right|\le \delta _1 (t)=\gamma _1 H_1
(\rho (t))\delta _M +c_0 ,
 \end{array}\end{equation}

Similarly, noting (9), (13), (39) and (41), we establish that
\begin{equation}
\begin{array}{l}
 \left| {\alpha _i^{(q)} -\alpha _i } \right|\le \delta _i (t)=\gamma _i H_i
(\rho (t))\left( {\delta _M +\delta _{i-1} (t)} \right)+c_0 , \\
 ~~~~~~~~~~~~~~~~~~~~~~~~~~~~~~~~~~~~~i=2,...,n \\
 \end{array}
\end{equation}

Substituting (42) into (38) yields
\begin{equation}
\dot {V}_n \le \left| {e_n } \right|\left( {H_n (\rho (t))-\gamma _n H_n
(\rho (t))\left| {e_n } \right|-c_n \left| {e_n } \right|^{N_n }+\delta _n
(t)} \right)
\end{equation}
on $\Omega $.

Noting $H_i (\rho (t))\ge H_i^\ast $ and invoking the definition of $p_i $,
it follows from the third inequality of (39) that
\begin{equation}
\begin{array}{l}
 H_i (\rho (t))+\gamma _i H_i (\rho (t))\left( {\delta _M +\delta _{i-1} }
\right)+\varepsilon _{i+1} +\delta _M+c_0 -c_i p_i^{N_i } \\
 \le \gamma _i H_i (\rho (t))\left( {\varepsilon _i +\delta _M +\delta
_{i-1} } \right)\le \gamma _i H_i (\rho (t))p_i \\
 \end{array}
\end{equation}

Using (44) and noting $p_{i+1} =\gamma _i H_i (\rho (t))\left( {\delta _M
+\delta _{i-1} } \right)+c_0+\varepsilon _{i+1} +\delta _M $ and $\delta _n =\gamma _n H_n (\rho (t))\left( {\delta _M
+\delta _{n-1} } \right) +c_0 $, one obtain
\begin{equation}
\left\{ {\begin{array}{l}
 H_i (\rho (t))\!-\!\gamma _i H_i (\rho (t))p_i \!-\!c_i p_i^{N_i } \!+\!p_{i+1}
\!<\!0,i=1,...,n-1 \\
 H_n (\rho (t))-\gamma _n H_n (\rho (t))p_n -c_n p_n^{N_n } +\delta _n <0 \\
 \end{array}} \right.
\end{equation}

It follows from (27) and (45) that, if $\left| {e_1 } \right|\ge p_1 $, then
$\dot {V}_1 \le 0$. Similarly, it follows from  (33), (43) and (45) that, if
$\left| {e_i } \right|\ge p_i $, then $\dot {V}_i \le 0$ for $i=2,...,n$.
Thus, we have
\begin{equation}
\left| {e_i } \right|\le p_i,~i=1,...,n
\end{equation}
which implies $e(t)\in \Omega $ for $\forall t\ge 0$, i.e., $\Omega $ is an
invariant set for $e(t)$. Then, it is not difficult to conclude the
boundedness of all the closed-loop signals.

From (46), it can be observed that
\begin{equation}
\left| {x_1 (t)-\rho (t)x_1 (0)} \right|\le p_1
\end{equation}
which further implies
\begin{equation}
\rho (t)x_1 (0)-\delta _M -\varepsilon _1 \le x_1 (t)\le \rho (t)x_1
(0)+\delta _M +\varepsilon _1
\end{equation}
This completes the proof.

\section{BFPPC of general nonlinear systems}
In this section, we will generalize the  above method to design a tracking controller for the following nonlinear systems
\begin{equation}
\left\{ {\begin{array}{l}
 \dot {x}_i =f_i (\bar {x}_i )+g_i (\bar {x}_i )x_{i+1} ,\quad i=1,...,n-1
\\
 \dot {x}_n =f_n (\bar {x}_n )+g_n (\bar {x}_n )u \\
 y=x_1 \\
 \end{array}} \right.
\end{equation}
where $f_i (\cdot )$ and $g_i (\cdot )$ are uncertain system nonlinearities,
$i=1,...,n$.
The control objective is to design a controller $u$ such that system output $y$ can track a given trajectory  $y_d $, and the tracking error is confined to a prescribed area.

\textit{Assumption 3}: $f_i (\cdot )$ and $g_i (\cdot )$ are continuous functions
satisfying $\left| {f_i (x_1 ,...,x_i )} \right|\le f_i^\ast (x_1 ,...,x_i
)$ and $0<g_m \le g_i (x_1 ,...,x_i )\le g_i^\ast (x_1 ,...,x_i )$ where
$f_i^\ast (\cdot )$ and $g_i^\ast (\cdot )$ are known functions for $i=1,...,n$. $g_m $ is a known constant.

\textit{Assumption 4}: The desired trajectory $y_d $ is continuously differentiable, and
$y_d $ and $\dot {y}_d $ are bounded, i.e. $\left| {y_d } \right|\le Y_0 $,
$\left| {\dot {y}_d } \right|\le Y_1 $ with $Y_0 $ and $Y_1 $ being known
positive constants.

\subsection{BFPPC controller}

Introduce the following error variables and change of coordinates
\begin{equation}
e_1 =x_1 -y_d -\rho _1 (t)(x_1 (0)-y_d (0))
\end{equation}
\begin{equation}
e_i =x_i -\alpha _{i-1} -\rho _i (t)x_i (0),\quad i=2,...,n
\end{equation}
where $\rho _i (t), i=1,...,n,$ is a continuously differentiable function satisfying 1) $\rho _i (0)=1$, $\rho _i (t)\ge 0$ and $\dot {\rho }_i (t)\le 0$ for
$t\ge 0$; 2) $\rho _i (t)$ and $\dot {\rho }_i (t)$ are bounded, namely,
$\left| {\rho _i (t)} \right|\le \rho _{M,i} $ and $\left| {\dot {\rho }_i
(t)} \right|\le \rho _{M,0} $ with $\rho _{M,i} >0,i=0,1,...,n$ being known
constants. $\rho _i (t)$ is the performance function for tracking error $e_i
$ since $\left| {e_i } \right|\le p_i $ will hold for $t\ge 0$, with $p_i
>0$ being an arbitrary small constant. Therefore, by noting that the
convergence rate of $\left| {x_1 -y_d } \right|$ can be determined by $\rho
_1 (t)$ in view of $\left| {e_1 } \right|\le p_1 $, we can design the
performance function for tracking error so as to achieved desired
convergence rate. For example, $\rho _i (t)$ can be designed as follows
\begin{equation}
\rho _i (t)\mbox{=}\left\{ {\begin{array}{l}
 \frac{1}{2}\left( {1+\cos \left( {\frac{t}{t_s }} \right)}
\right),\quad t<\pi t_s \\
 0,\quad t\ge \pi t_s \\
 \end{array}} \right.
\end{equation}
or
\begin{equation}
\rho _i (t)\mbox{=}\left( {1-\rho _{i,1} } \right)e^{-\rho _{i,0} t}+\rho
_{i,1}
\end{equation}
where $t_s $, $\rho _{i,0} $ and $\rho _{i,1}<1 $ are arbitrary positive constants.

The BFPPC controller $u$ and virtual controllers $\alpha _i $ are
designed as follows
\begin{equation}
\begin{array}{l}
 \alpha _i =-k_i e_i -M_i \tanh \left(
{\frac{M_ie_i }{\varepsilon _i }} \right) -c_i e_i^{N_i } , ~~i=2,...,n-1 \\
 \end{array}
\end{equation}
\begin{equation}\begin{array}{l}
u=-k_n e_n -M_n \tanh \left(
{\frac{M_ne_n }{\varepsilon _n }} \right) -c_n e_n^{N_n }
 \end{array}\end{equation}
where $k_i $, $c_i $, $M_i $ and $\varepsilon _i $ are positive design
parameters, with $N_i >1$ being arbitrary odd integers.

\subsection{Main results}
For system (49) with our designed BFPPC controller, we have the following
results.

\textbf{\textit{Theorem 2}}: Consider the closed-loop system consisting of
uncertain nonlinear system (49) satisfying Assumption 3-4, the virtual
control signals (54) and the actual controller (55). Then, there exist
$k_i $, $c_i $, $\varepsilon _i $, $N_i $ and $M_i $ such that the
following properties hold:

1). The GBPPC problem of system (49) is solved.

2). The tracking error satisfies the prescribed performance, specially,
$\rho _1^\ast (t)-p_1 \le x_1 -y_d \le \rho _1^\ast (t)+p_1 $ where $\rho
_1^\ast (t)=\rho _1 (t)(x_1 (0)-y_d (0))$.

\textit{Proof}: Define a compact set
\begin{equation}
\Omega =\left\{ {(e_1 ,...,e_n )\left| {\left| {e_i } \right|\le p_i }
\right.,i=1,...,n} \right\}
\end{equation}
where $p_i>0$ are arbitrary small constants.

Denote $e=[e_1 ,...,e_n ]^T$. It can be easily verified that $e(0)\in
\Omega $ by noting $e_1 (0)=e_2 (0)=\cdots =e_n (0)=0$.

In the following, we will prove that $e(t)\in \Omega $ for $\forall t\ge 0$,
i.e., $\Omega $ is an invariant set for $e(t)$.

Step $i\;(i=1,...,n-1)$: Consider the following Lyapunov function
\begin{equation}
\label{eq4}
V_i =\frac{1}{2}e_i^2
\end{equation}

Noting (49)-(51), the time derivative of $e_i $ is
\begin{equation}
\dot {e}_i \!=\!g_i (\bar {x}_i )\left( {e_{i+1} \!+\!\rho _{i+1} (t)x_{i+1} (0)\!+\!\alpha
_i } \right)\!+\!f_i (\bar {x}_i )\!-\dot {\alpha }_{i-1} \!-\!\dot {\rho }_i (t)x_i
(0)
\end{equation}
where $\dot {\alpha }_0 =\dot {y}_d+\dot {\rho }_1 (t)y_d
(0) $ and $e_{n+1} =x_{n+1} (0)=0$. And denote $a_0=|x_1(0)-y_d(0)|$.

Similarly as Theorem 1, there exist $W$-functions $\varphi _{i,1} (\cdot )$ and
$\varphi _{i,2} (\cdot )$ satisfying
\begin{equation}
\left| {f_i (\bar {x}_i )} \right|\le \varphi _{i,1} (Z_i^T )
\end{equation}
\begin{equation}
\left| {g_i (\bar {x}_i )} \right|\le \varphi _{i,2} (Z_i^T )
\end{equation}
where $Z_i =[\rho _1 (t),...,\rho _i (t),\left| {e_1 } \right|,...,\left|
{e_i } \right|,\left| {y_d } \right|,a_0 ,\left| {x_2 (0)}
\right|,...,\\\left| {x_i (0)} \right|]^T$. Thus, there exists a $W$-function
$F_i^0 (\cdot )$ such that
\begin{equation}
\begin{array}{l}
 F_i^0 \Big(\rho _1 (t),...,\rho _i (t),\left| {\dot {\rho }_1 (t)}
\right|,...,\left| {\dot {\rho }_i (t)} \right|,\left| {e_1 }
\right|,...,\left| {e_{i+1} } \right|,
\\~~~~~\left| {y_d } \right|,|\dot{y}_d|,a_0 ,\left| {x_2
(0)} \right|,...,\left| {x_{i+1} (0)} \right|\Big) \\
 \ge \frac{1}{g_m }\big(\varphi _{i,1} (Z_i^T )+\varphi _{i,2} (Z_i^T )\left( {\left|
{e_{i+1} } \right|+\left| {\rho _{i+1} (t)x_{i+1} (0)} \right|} \right)\!+\!\left|
{\dot {\alpha }_{i-1} } \right| \\
 \quad~~~~~~~ +\left| {\dot {\rho }_i (t)x_i (0)} \right|\big) \\
 \end{array}
\end{equation}

Define constant $F_i^\ast$ as
\begin{equation}
\begin{array}{l}
 F_i^\ast  =F_i^0 \Big(\rho _{M,1},...,\rho _{M,i},\rho _{M,0},...,\rho _{M,0},p_1 ,...,p_{i+1} ,\\ ~~~~~~~~~~~~~Y_0,Y_1,a_0 ,\left| {x_2 (0)} \right|,...,\left|
{x_{i+1} (0)} \right|\Big) \\
 \end{array}
\end{equation}

Therefore, it follows from (59)-(62) and the property of $W$-function that
\begin{equation}
\begin{array}{l}
 F_i^\ast \ge \frac{1}{g_m }\big(\left| {f_i (\bar {x}_i )} \right|\!+\left| {g_i
(\bar {x}_i )} \right|\left( {\left| {e_{i+1} } \right|+\left| {\rho _{i+1}
(t)x_{i+1} (0)} \right|} \right)\!+\!\left| {\dot {\alpha }_{i-1} } \right| \\
 \quad ~~~~~~~~~~~~+\left| {\dot {\rho }_i (t)x_i (0)} \right|\big) \\
 \end{array}
\end{equation}
on $\Omega$.

Similarly, using (57), (58) and (63) and employing $ -M_i e_i\tanh \left(
{\frac{M_ie_i }{\varepsilon _i }} \right)\le -M_i|e_i |+0.3\varepsilon _i$ according to Lemma 2 in [15], the time derivative of $V_i $ satisfies
\begin{equation}
\begin{array}{l}
 \dot {V}_i \le g_i (\bar {x}_i )e_i \alpha _i +g_m\left| {e_i }
\right|F_i^\ast \\
 \le -k_i g_m e_i^2 -g_m \left| {e_i } \right|M_i -c_i g_m e_i^{N_i +1} \\
 \quad +0.3\varepsilon _i g_m+g_m\left| {e_i } \right|F_i^\ast \\
 \end{array}
\end{equation}
on $\Omega $.

Choose $k_i $, $M_i $, $c_i $ and $\varepsilon _i $ to satisfy
\begin{equation}
k_i p_i +M_i +c_i p_i^{N_i} \ge F_i^\ast +0.3\frac{1}{p_i
}\varepsilon _i
\end{equation}

There always exist parameters satisfying (65) by increasing either of
$k_i $, $M_i $ or $c_i $, since the righthand of (65) is a known constant.

It follows from (64) and (65) that $\dot {V}_i \le 0$ when $\left| {e_i }
\right|\ge p_i $. Thus, we have
\begin{equation}
\left| {e_i } \right|\le p_i
\end{equation}

Step $n$: Consider the following Lyapunov function
\begin{equation}
V_n =\frac{1}{2}e_n^2
\end{equation}

The time derivative of $e_n $ is
\begin{equation}
\dot {e}_n =f_n (\bar {x}_n )+u-\dot {\alpha }_{n-1} -\dot {\rho }_n (t)x_n
(0)
\end{equation}

By some similar derivations as the former steps, it can be deduced that
there exist $W$-functions $F_n^0 (\cdot )$ and constant $F_n^\ast $ such
that
\begin{equation}
\begin{array}{l}
 F_n^\ast  =F_n^0 \Big(\rho _{M,1},...,\rho _{M,n},\rho _{M,0},...,\rho _{M,0},p_1 ,...,p_{n} ,\\ ~~~~~~~~~~~~~Y_0,Y_1,a_0 ,\left| {x_2 (0)} \right|,...,\left|
{x_n (0)} \right|\Big) \\
 \end{array}
\end{equation}
and
\begin{equation}
\begin{array}{l}
 \dot {V}_n \le g_m e_n u+\left| {e_n } \right|g_m F_n^\ast  \\
 \le -k_n g_m e_n^2 -g_m \left| {e_n } \right|M_n-c_n g_m e_n^{N_n +1}  \\
 ~~~+0.3g_m \varepsilon _n+\left| {e_n } \right|g_m F_n^\ast \\
 \end{array}
\end{equation}
on $\Omega $.

Choose $k_n $, $M_n$, $c_n $ and $\varepsilon _n $ to satisfy
\begin{equation}
k_n p_n +M_n +c_n p_n^{N_n } \ge F_n^\ast +0.3\frac{1}{p_n
}\varepsilon _n
\end{equation}

It is easily known that (71) can be satisfied by increasing either of $k_n $, $M_n$ or $c_n $.

It follows from (70) and (71) that $\dot {V}_n \le 0$ when $\left| {e_n }
\right|\ge p_n $. Thus, we have
\begin{equation}
\left| {e_n } \right|\le p_n
\end{equation}

From (66) and (72), it can be seen that $e(t)\in \Omega $ for $\forall
t\ge 0$, namely, $\Omega $ is an invariant set for $e(t)$. Using (50) and
(66), we can further have
\begin{equation}
\left| {x_1 -y_d -\rho _1 (t)(x_1 (0)-y_d (0))} \right|\le p_1
\end{equation}
which implies
\begin{equation}
\rho _1^\ast (t)-p_1 \le x_1 -y_d \le \rho _1^\ast (t)+p_1
\end{equation}
where $\rho _1^\ast (t)=\rho _1 (t)(x_1 (0)-y_d (0))$. This completes the
proof.

\textit{Remark 3}: The invariant set $\Omega $ plays a key role in reducing the
complexity of virtual and actual controller. Noting $F_i^0 (\cdot )$ is an
increasing function with respect to $\rho _1 (t),...,\rho _i (t),\left|
{\dot {\rho }_1 (t)} \right|,...,\left| {\dot {\rho }_i (t)} \right|,\left|
{e_1 } \right|,...,\left| {e_i } \right|,\left| {y_d } \right|,|\dot{y}_d|$, we use its
maximum, $F_i^\ast$, since
all the variables has maximum on the introduced invariant set $\Omega $,
i.e., $\left| {e_i } \right|\le p_i $. It can be seen that the time-varying
terms $\left| {\dot {\rho }_1 (t)} \right|,...,\left| {\dot {\rho }_i (t)}
\right|,\left| {e_1 } \right|,...,\left| {e_{i-1} } \right|,\left| {y_d }
\right|,|\dot{y}_d|$ do not appear in $\alpha _i $ by using their maximums on $\Omega
 $, and thus $\alpha _i $ is only a function of $e_i $ and therefore it is a low-complexity controller.

\section{Optimizing BFPPC controller parameters}
Since we enlarge some terms in $W$-function in choosing the controller
parameters, therefore, the controller parameters are always conservative as they are usually larger or
smaller than they should to be. In this part, we set a switching mechanism for
the selection of controller parameters so as to find out more suitable
controller parameters.

Select a series of constants $\left\{ {p_{i,m} } \right\}$ such that
$0<p_{i,1} <p_{i,2} <\cdots <p_{i,K} $ for $i=1,...,n$ and $m=1,...,K$,
where $K$ is a positive integer specified by designer and it represents the
maximum of switching times.

Denote the switching time instant as $t_{\sigma (t)} $ with $\sigma (t)$
defined as follows
\begin{equation}
\sigma (t)=\left\{ {\begin{array}{l}
 m+1,\quad otherwise \\
 m,\quad if\;\left| {e_i } \right|\le p_{i,m} ,for\;all\;i\in \{1,...,n\} \\
 \end{array}} \right.
\end{equation}
which implies $\left| {e_i } \right|\le p_{i,m} $ for $0\le t\le t_m $,
where $t_0 \mbox{=}0$.

\textbf{\textit{Theorem 3}}: Consider the nonlinear system (49) with
Assumption 3-4, the virtual controllers (54), the actual controller (55). The
controller parameters are chosen as $k_i =k_{i,\sigma (t)} $, $M_i
=M_{i,\sigma (t)} $, $c_i =c_{i,\sigma (t)} $, $N_i =N_{i,\sigma (t)} $, for
$t\in [t_{\sigma (t)-1} ,t_{\sigma (t)} )$, where $k_{i,\sigma (t)} $,
$M_{i,\sigma (t)} $, $c_{i,\sigma (t)} $ and $N_{i,\sigma (t)} ,i=1,...,n$
for $\sigma (t)\le K-1$ are chosen arbitrarily, while $k_{i,K} $, $M_{i,K}
$, $c_{i,K} $ and $N_{i,K} ,i=1,...,n$ are chosen to satisfy (65) and (71), with
\begin{equation}
\begin{array}{l}
 F_i^\ast  =F_i^0 \Big(\rho _{M,1},...,\rho _{M,i},\rho _{M,0},...,\rho _{M,0},p_{1,K} ,...,p_{i+1,K},\\ ~~~~~~~~~~~~~Y_0,Y_1,a_0 ,\left| {x_2 (0)} \right|,...,\left|
{x_{i+1} (0)} \right|\Big) \\
 \end{array}
\end{equation}
\begin{equation}
\begin{array}{l}
 F_n^\ast  =F_n^0 \Big(\rho _{M,1},...,\rho _{M,i},\rho _{M,0},...,\rho _{M,0},p_{1,K} ,...,p_{n,K},\\ ~~~~~~~~~~~~~Y_0,Y_1,a_0 ,\left| {x_2 (0)} \right|,...,\left|
{x_{i+1} (0)} \right|\Big) \\
 \end{array}
\end{equation}
Then, global boundedness of all the system signals is guaranteed.

\textit{Proof}: By noting $\left| {e_i } \right|\le p_{i,K} $ for $0\le t\le t_K $,
it is not difficult to prove Theorem 3 by following the similar way as the proof of
Theorem 2. Thus the proof is omitted.

\textit{Remark 4}: Theorem 3 is introduced not only to guarantee the global stability of system
but also to seek more preferable controller parameters. The controller parameters $k_{i,\sigma (t)} $,
$M_{i,\sigma (t)} $, $c_{i,\sigma (t)} $ and $N_{i,\sigma (t)} ,i=1,...,n$
for $\sigma (t)\le K-1$ can be chosen arbitrarily, which suggests the control objective may be achieved by the
controller parameters determined freely by experienced designer while the
system stability is also guaranteed in this method.

\textit{Remark 5}: Theorem 3 suggests that the prescribed performance may be satisfied
at $\sigma (t)\le K-1$, before $W$-functions are used to select the
controller parameters, which may result in better control action while
prescribed performance are also guaranteed, since the selection of
controller parameters may combine the experience of designer.

\textit{Remark 6}: The essential difference from all the existing prescribed
performance control is, that, the controller in this paper does not contain
barrier functions, as shown as follows
\begin{equation}
\begin{array}{l}
 \alpha _i =\!-k_i e_i \!-M_i \tanh \left( {\frac{M_i e_i }{\varepsilon _i }}
\right)-c_i e_i^{N_i } ,~ i=1,...,n-1 \\
 \end{array}
\end{equation}
\begin{equation}
u=-k_n e_n -M_n \tanh \left( {\frac{M_n e_n }{\varepsilon _n }} \right)-c_n
e_n^{N_n }
\end{equation}
All the other prescribed performance control methods are mostly derived from [21], which controller are, for example, given as follows
\begin{equation}
\alpha _i =-k_i \ln \left( {\frac{1+\frac{e_i }{\rho _i }}{1-\frac{e_i
}{\rho _i }}} \right),i=1,...,n-1
\end{equation}
\begin{equation}
u=-k_n \ln \left( {\frac{1+\frac{e_n }{\rho _n }}{1-\frac{e_n }{\rho _n }}}
\right)
\end{equation}
Comparing with these controllers, the reasons why we design the controller
construction as (78) and (79) are that: \textbf{(1)} Firstly, barrier functions, such as $\ln
(\cdot )$, may easily run to an error or singularity in practice. \textbf{(2)} Secondly,
expanding (80) and (81) by using Taylor series as follows
\begin{equation}
\alpha _i =2k_i \left( {-\frac{e_i }{\rho _i }-\frac{1}{3}\left( {\frac{e_i
}{\rho _i }} \right)^3-\frac{1}{5}\left( {\frac{e_i }{\rho _i }}
\right)^5-\cdots } \right),i=1,...,n-1
\end{equation}
\begin{equation}
u=2k_n \left( {-\frac{e_n }{\rho _n }-\frac{1}{3}\left( {\frac{e_n }{\rho _n
}} \right)^3-\frac{1}{5}\left( {\frac{e_n }{\rho _n }} \right)^5-\cdots }
\right)
\end{equation}
It can be seen that our controller construction preserves some high-order
terms of (82) and (83), such as $c_i e_i^{N_i } $, which makes our controller preserve
some quality of traditional PPC, and discards residual high-order terms,
which makes our controller avoid singularity facing the case of $e_i $ being
discontinuous. \textbf{(3)} Thirdly, the term, $M_i \tanh \left( {\frac{M_i e_i }{\varepsilon
_i }} \right)$, is used to avoid the controller action to decreasing
dramatically when $e_i $ is small.

These characteristics ensure our controller to have better control
performance than traditional PPC and other normal controllers.

\section{Simulation results}

In this section, two simulation examples are presented to demonstrate the
advantages of BFPPC method. Consider the following system with quantized
states
\begin{equation}
\left\{ {\begin{array}{l}
 \dot {x}_1 =x_1^2 -\sin x_1 +x_2 \\
 \dot {x}_2 =x_1 x_2^2 +u \\
 \end{array}} \right.
\end{equation}

For the purpose of simulation, let $x_1 (0)=1$, $x_2 (0)=0$, and use uniform
quantizer (3) with $l_0 =0.1$. According to our BFPPC method, the $W$-function should be chosen to satisfy
\begin{equation}
\begin{array}{l}
{H_1}(\rho (t)) \ge {\left( {{p_1} + \rho (t)\left( {\left| {x_1^{(q)}(0)} \right| + 0.1} \right)} \right)^2}\\
\quad \quad\quad\quad\quad  + 1 + \frac{1}{2}\left( {\left| {x_1^{(q)}(0)} \right| + 0.1} \right) + \left( {\left| {x_2^{(q)}(0)} \right| + 0.1} \right)
\end{array}
\end{equation}
\begin{equation}
\begin{array}{l}
{H_2}\left( {\rho (t)} \right) \ge \left| {{{\dot \alpha }_1}} \right| + \frac{1}{2}\left( {\left| {x_2^{(q)}(0)} \right| + 0.1} \right)\\
 + \!\left( {{p_1}\! +\! \rho (t)\left( {\left| {x_1^{(q)}(0)} \right| \!+ \!0.1} \right)} \right){\left( {{p_2} \!+\! \left| {{\alpha _1}} \right| \!+ \!\rho (t)\left( {\left| {x_2^{(q)}(0)} \right| + 0.1} \right)} \right)^2}
\end{array}
\end{equation}
Then, noting $|\rho (t)|\le 1$, we can choose the design parameters and functions
as $p_1 =\delta _M +0.05$, $p_2 =2+\delta _M +1$, $H_1 (\rho (t))=5$,
$\gamma _1 =4$, $c_1 =0.1$, $N_1 =3$, $H_2 (\rho (t))=10$, $\gamma _2
=0.4$, $c_2 =1.5$, $N_2 =3$. Set performance function $\rho(t)$ as (8) with $t_s =1$. According to Theorem 1, the virtual and actual controller are designed as
follows
\begin{equation}\begin{array}{l}
\alpha _1 =-20e_1^{(q)} -0.1\left[ {e_1^{(q)} } \right]^3,
\end{array}\end{equation}
\begin{equation}\begin{array}{l}
u=-4e_2^{(q)} -1.5\left[ {e_2^{(q)} } \right]^3,
\end{array}\end{equation}
The simulation results are depicted as Figs. 2-5. It can be seen that from
these results that, though system nonlinearities do not satisfy global
Lipschitz continuity condition, the prescribed performance for system
output is achieved with quantized signals which makes the barrier
functions-based controller unavailable.

\begin{figure}[h!]
    \includegraphics[width=2.2in,height=2.2in,clip,keepaspectratio]{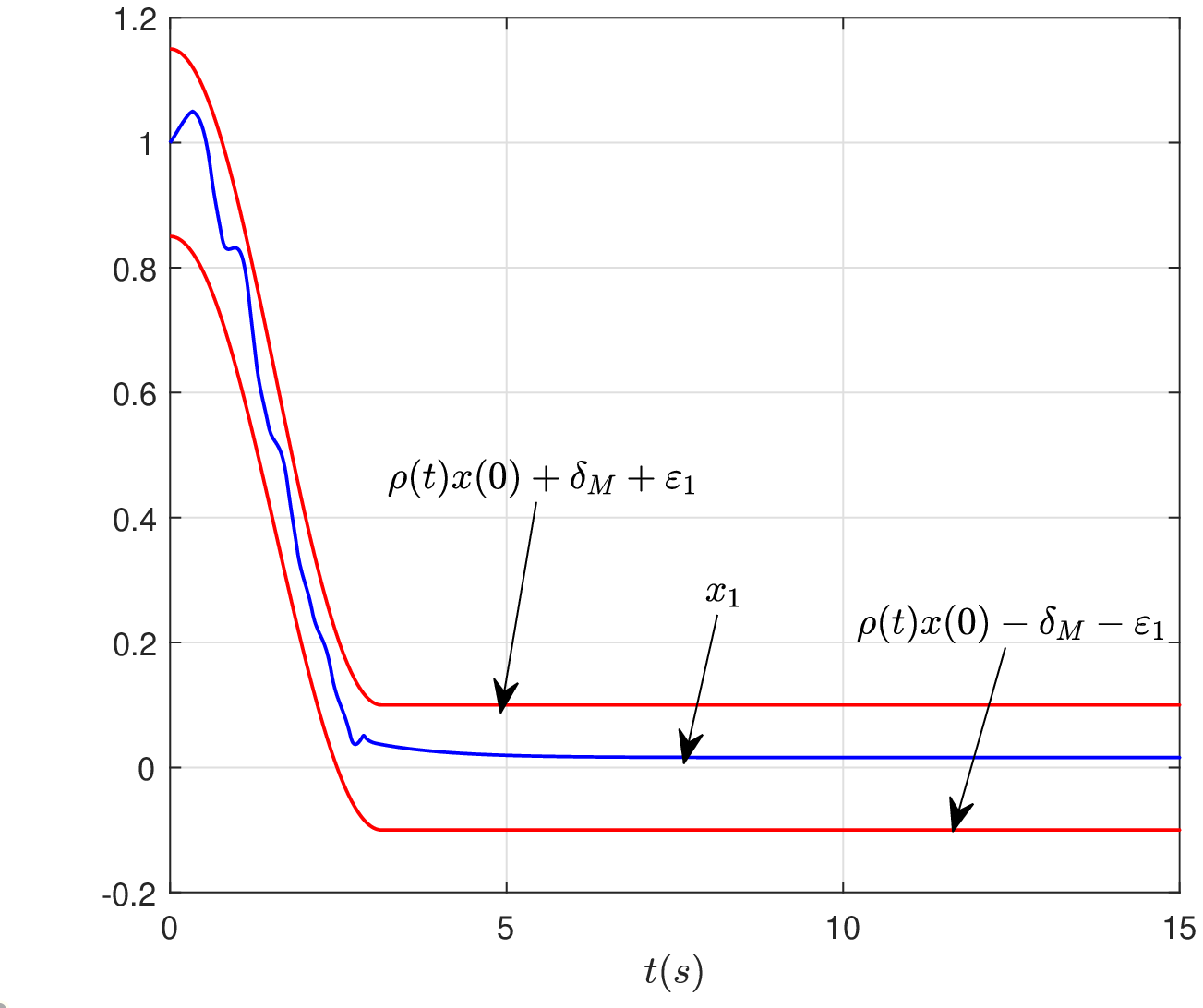}\\
  \raggedleft
  \caption{System output $y$ and its prespecified bounds}
\end{figure}
\begin{figure}[h!]
    \includegraphics[width=2.2in,height=2.2in,clip,keepaspectratio]{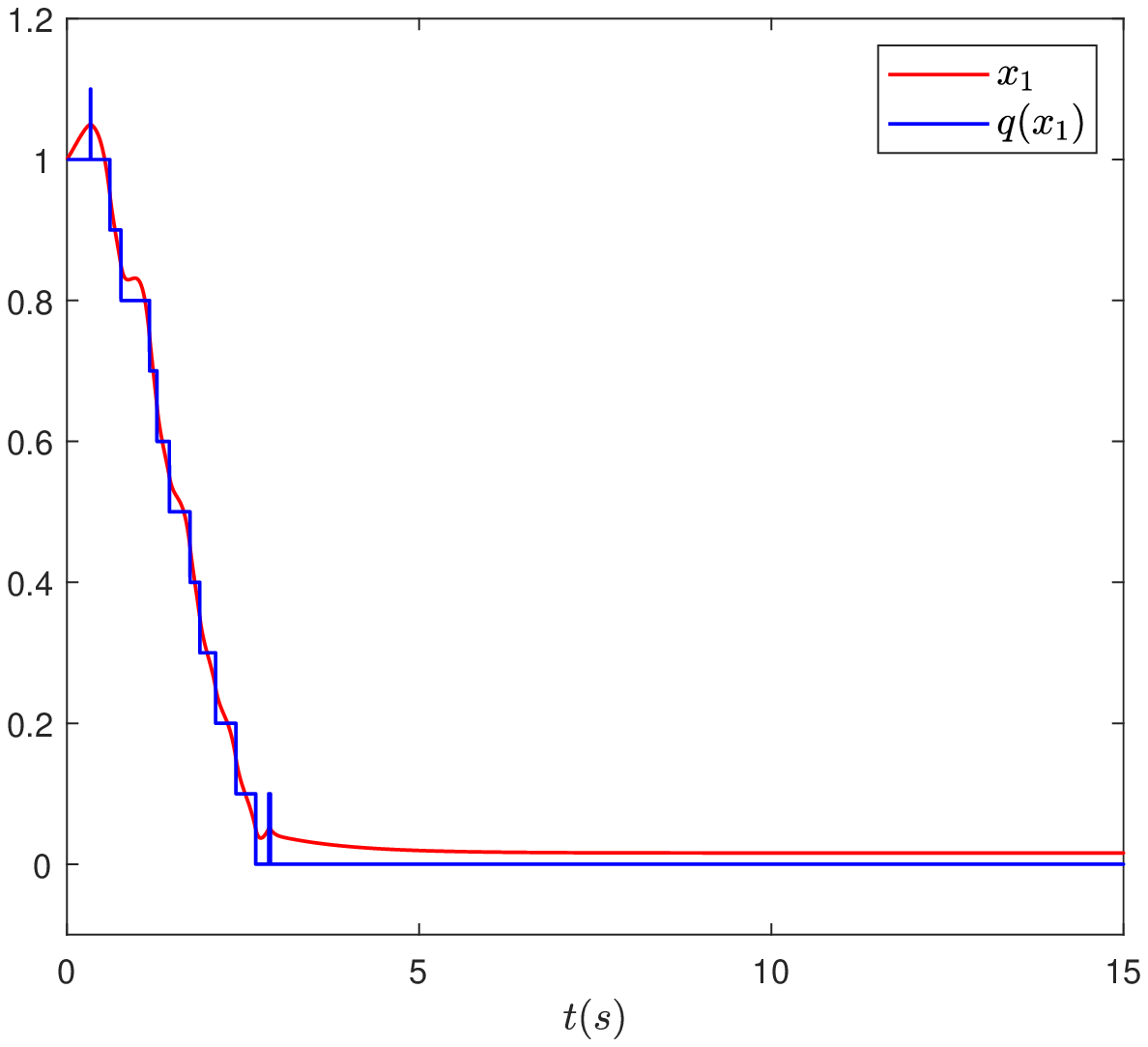}\\
  \raggedleft
  \caption{System state $x_1 $ and quantized signal $q(x_1 )$}
\end{figure}
\begin{figure}[h!]
    \includegraphics[width=2.2in,height=2.2in,clip,keepaspectratio]{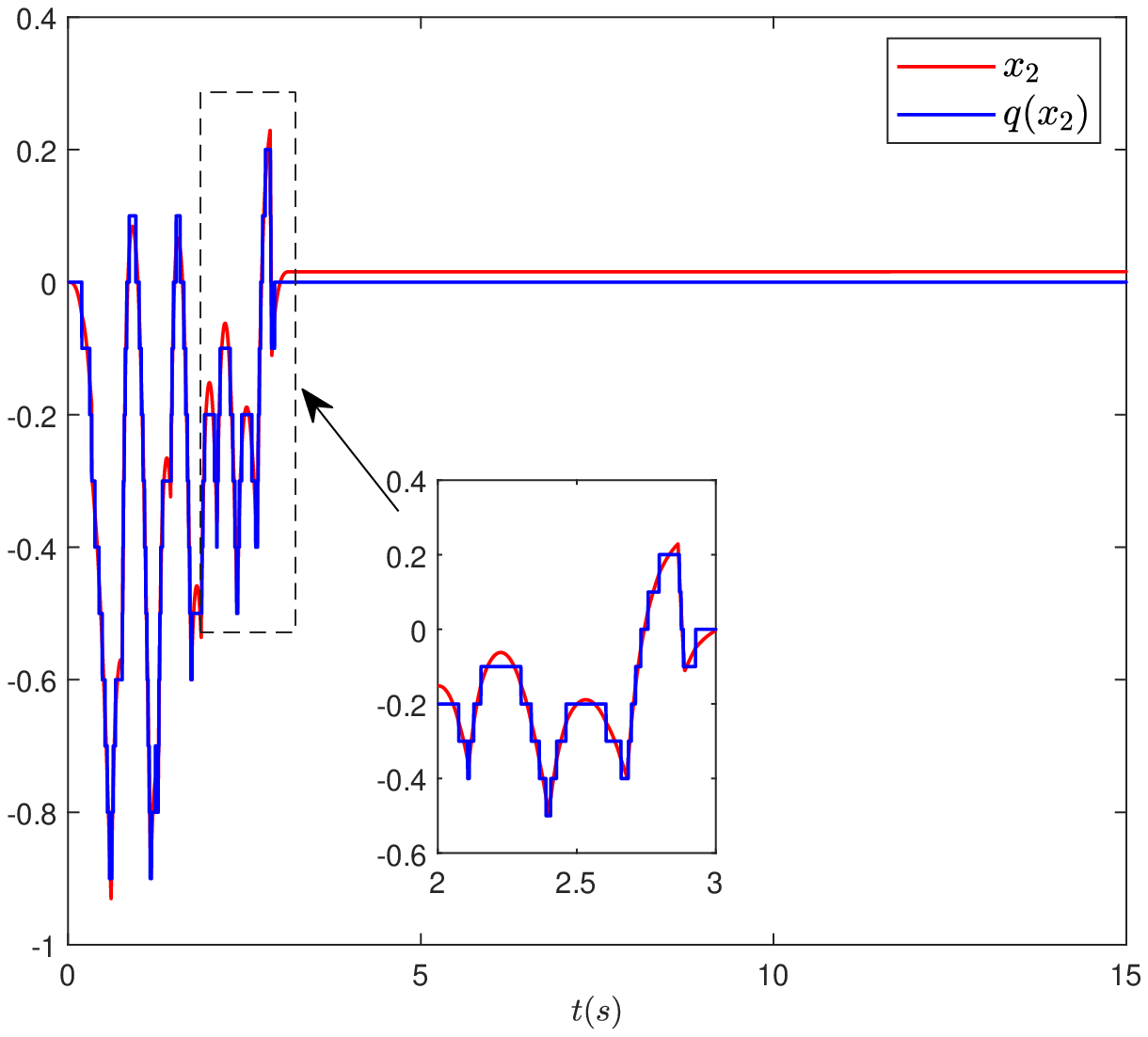}\\
  \raggedleft
  \caption{System state $x_2 $ and quantized signal $q(x_2 )$}
\end{figure}
\begin{figure}[h!]
    \includegraphics[width=2.2in,height=2.2in,clip,keepaspectratio]{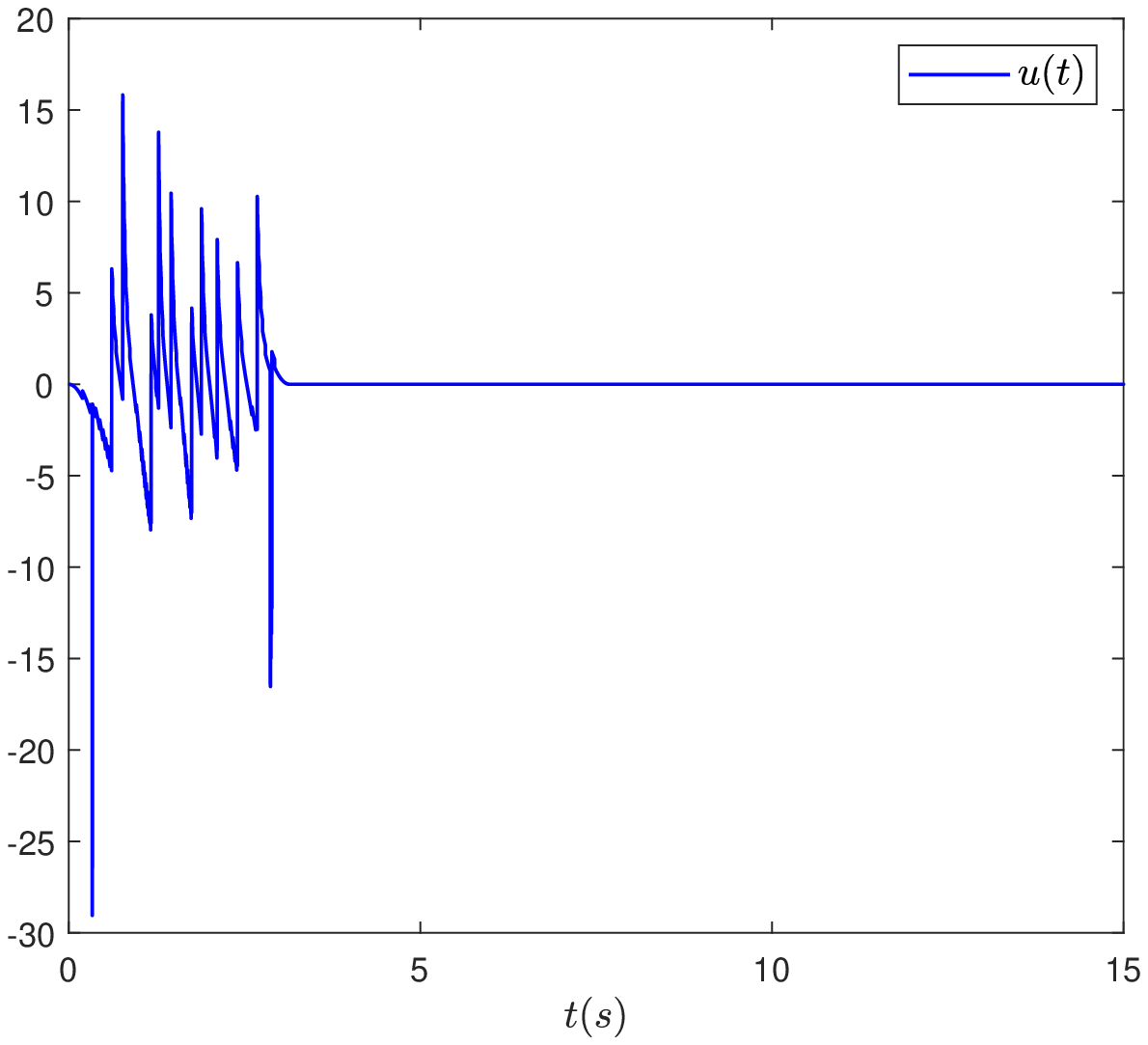}\\
  \raggedleft
  \caption{Control input $u$}
\end{figure}
\begin{figure}[h]
    \includegraphics[width=2.2in,height=2.2in,clip,keepaspectratio]{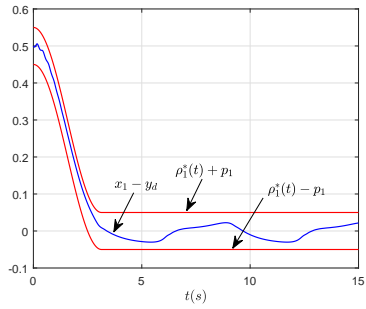}\\
  \raggedleft
  \caption{Tracking error and its prespecified bounds}
\end{figure}
\begin{figure}[h!]
    \includegraphics[width=2.2in,height=2.2in,clip,keepaspectratio]{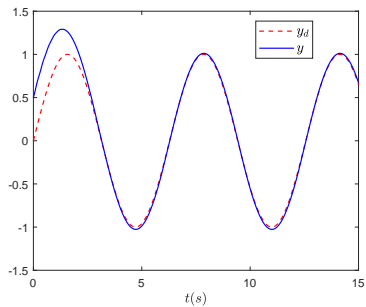}\\
  \raggedleft
  \caption{System output $y$ and desired signal $y_d $}
\end{figure}
\begin{figure}[h!]
    \includegraphics[width=2.2in,height=2.2in,clip,keepaspectratio]{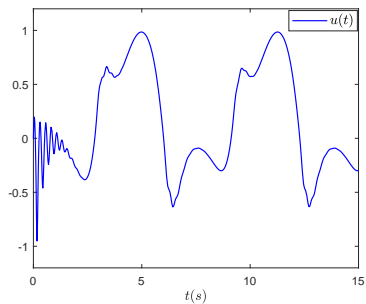}\\
  \raggedleft
  \caption{Control input $u$}
\end{figure}

To further show the advantage of BFPPC, consider the following uncertain
nonlinear system:
\begin{equation}
\left\{ {\begin{array}{l}
 \dot {x}_1 =x_1 +x_1 e^{-0.5x_1 }+\left( {1+\sin x_1^2 } \right)x_2 \\
 \dot {x}_2 =x_1 \sin x_2 +x_1 x_2^2 +(3+\cos x_1 )u \\
 \end{array}} \right.
\end{equation}
where $y_d =\sin t, f_1 (x_1 )=x_1 +x_1 e^{-0.5x_1 }$, $g_1 (x_1 )=1+\sin
x_1^2 $, $f_2 (\bar {x}_2 )=x_1 \sin x_2 +x_1 x_2^2 $, $g_2 (\bar {x}_2
)=3+\cos x_1 $, and $f_1^\ast (x_1 )=\left| {x_1 } \right|+\left| {x_1 }
\right|e^{0.5\left| {x_1 } \right|}$, $g_1^\ast (x_1 )=2$, $g_m =1$,
$f_2^\ast (\bar {x}_2 )=\left| {x_1 } \right|+\left| {x_1 } \right|x_2^2 $,
$g_2^\ast (\bar {x}_2 )=4$. Let $x_1 (0)=0.5$, $x_2 (0)=0$. Choose $\rho_1(t)$ and $\rho_2(t)$ as (52) with $t_s =1$. According to
Theorem 3, the BFPPC controller are constructed as follows
\begin{equation}
\alpha _1 =-k_{1,\sigma (t)} e_1 -M_{1,\sigma (t)} \tanh \left(
{\frac{M_{1,\sigma (t)} e_1 }{0.5}} \right)-c_{1,\sigma (t)}
e_1^{N_{1,\sigma (t)} } ,
\end{equation}
\begin{equation}
u=-k_{2,\sigma (t)} e_2 -M_{2,\sigma (t)} \tanh \left( {\frac{M_{2,\sigma
(t)} e_2 }{0.5}} \right)-c_{2,\sigma (t)} e_2^{N_{2,\sigma (t)} } ,
\end{equation}
Choose $K=2$, $p_{1,1} =0.04$, $p_{2,1} =1$, $p_{1,2} =0.05$, $p_{2,2} =2$,
$k_{1,1} =2$, $c_{1,1} =0.1$, $M_{1,1} =6$, $k_{2,1} =1$, $c_{2,1} =2$,
$M_{2,1} =0.1$, $N_{1,1} =N_{2,1} =3$, while $k_{1,2} =k_{2,2} =2$, $N_{1,2}
=3$, $N_{2,2} =5$, $M_{i,2} $ and $c_{i,2} $, $i=1,2$, are any constants
that satisfying (65) and (71) with $F_1^\ast =F_1^0 (1,1,0.05,2,1,1,0.5,0)$ and
$F_2^\ast =F_2^0 (1,1,1,1,0.05,2,1,1,0.5,0)$. The simulation results are
depicted as Figs. 6-8. From Fig. 6, it can be seen that the prescribed
performance for tracking error is achieved by the proposed controller.

\section{Conclusion}
This paper proposes BFPPC method based on the an invariant set and $W$-functions. By using the maximums of some variables or functions on the invariant set, the designed controller is simplified such that repeated differentiation of virtual controls is avoided and prescribed performance is also guaranteed. This novel method is extended to solve the control problem of nonlinear system with state quantization. By virtue of BFPPC method, the global Lipschitz continuity condition is cancelled and global bounded of all the closed-loop signals is proved. Simulation results demonstrate the effectiveness of our method.

\end{document}